# Acid-mediated tumor invasion: how does vasculature affect the growth characteristics?

**B. S. Govindan** [1], **W. B. Spillman, Jr.** [1, 2], **J. L. Robertson** [3] and **W. R. Huckle** [4]

(1) Virginia Tech Applied Biosciences Center
(2) Department of Physics
(3) Center for Comparative Oncology at Virginia Tech
(4) Department of Biomedical Sciences and Pathobiology,

*Virginia Polytechnic Institute and State University*
*Blacksburg, VA 24061, USA.*

## Abstract

We study the growth of an implanted avascular tumor in rats, in two-dimensions, based on a model where the mechanism of invasion is centered on tumor-induced acidification of the micro-environment and consequent death of normal cells. The spatial distribution of the acid density around the tumor is found using mean-field analysis. By assuming that the viability of both normal and tumor cells falls sharply below certain threshold values of the local pH, we determine the conditions for the formation and radius of a necrotic core at the center, as a function of the tumor radius. We show mathematically that the mean micro-vessel density (MVD) plays a pivotal role in determining the growth characteristics of the tumor. When the MVD is sufficiently small, accumulation of excess acid inside the tumor leads to the formation of a necrotic core, which occupies a significant fraction of the total area in large tumors. However, necrosis is reduced when the mean MVD inside the tumor is larger than outside because of the more efficient removal of excess acid. At sufficiently high MVD, necrosis might be absent in the tumor, or confined to small regions mostly devoid of micro-vessels. Quantitative estimates of MVD for these different phases of growth are obtained, and verified using explicit cellular automaton simulations. Recent experimental studies on the correlation between necrosis and MVD support our main conclusions.

¶ **Corresponding author**: E-mail: wspillma@vt.edu, Fax : +1-540-231 6642

# 1. Introduction

It is now well accepted that several phenotypic traits of cancer, such as invasion and metastasis, may arise from the unique physiological environments created by tumors. Such environments may have features such as low pH, hypoxia and nutrient depletion. These micro-environmental changes adversely affect the normal cells in the vicinity of the tumor, leading to disruption of metabolism, degradation of protein synthesis and even damage to their DNA. Such micro-environmental stresses therefore either lead to destruction of normal cells which would be subsequently replaced by dividing tumor cells, or favor the evolution of subpopulations of more aggressive phenotypes which are better adapted to surviving in adverse conditions . Both of these factors contribute to the malignancy (tendency for invasion and metastasis) of the tumor (Rockwell *et. al.,* 2001).

Warburg's early investigations (Warburg, 1930) revealed that some tumor cells have a high rate of anaerobic glycolysis, even in the presence of ample oxygen, a feature resulting in the production of large amounts of lactic acid compared to normal tissue. This propensity for anaerobic metabolism may be important for tumor cell success, since, in many developing tumors, the supporting vasculature is often limiting, leading to an insufficient supply of oxygen to all regions of the tumor. Tumor cells can rapidly adapt to this hypoxic microenvironment by upregulating several genes, including those controlling anaerobic glycolysis. It must, however, be noted that the intra-cellular pH of tumor cells is close to neutral, or even slightly alkaline, so the excess H+ ions diffuse into the surrounding intercellular fluid, lowering its pH significantly. Although lactic acid has long been thought to be the primary source of acidity, recent experiments indicate that the low pH in the intercellular space around a tumor could also have its origin as excreted $CO_2$, which forms carbonic acid (Griffiths, et. al., 2001; Helmlinger et. al., 2002).

Gatenby and Gawlinski (1996) proposed that tumor invasion of a normal tissue is primarily due to the low intracellular pH in the space surrounding a tumor. According to this model, the excess acid produced by cancer cells and the consequent decrease in intercellular pH leads to degradation of vital biochemical processes of the normal cells, and ultimately to their death. Their model assumes that tumors grow primarily by filling the hypo-cellular gap, created by the death of normal cells, through cell division. Tumor cells are assumed to be more resistant to low pH levels than normal cells, and thus can survive in the more acidic microenvironment created by cell necrosis and degradation.

Invasive tumors secrete degradative proteases that function at neutral and acidic pH, to infiltrate surrounding tissue. The activity and expression of heparanases, acid cathepsins, matrix metalloproteinases, and plasminogen activators have been demonstrated both in tissue culture and in tumor biopsies (Chambers *et. al*., 1997). Such enzymes facilitate degradation of extracellular matrix, allowing adhesion and attachment of tumor cells, as well as removing critical tissue barriers (connective tissue) that can physically impede or retard the movement of cells. The presence of this enzyme-rich and degraded environment not only facilitates tumor cell movement, but is a key to sprouting and integration of new vasculature. A wide variety of enzyme inhibitors modulate the extent and speed of this process.

The importance of the micro-vascular network, particularly the micro-vessel density (MVD), as a modulator of tumor invasiveness has been a source of debate. It is known that the disruption of existing micro-vasculature by growing tumor cells, and consequent hypoxia, activates angiogenesis, leading to neovascularization of the tumor. However, in spite of the high overall vascularization, some regions inside the tumor may be poorly perfused. The likely explanations are that, either the MVD inside the tumor is very heterogeneous, and/or only a fraction of the newly formed micro-vessel network is functional (Bernsen et. al., 1995 p.64). Modeling the dynamics of physiological environment, tumor metabolism, vascularization and metastasis could be very helpful for developing new paradigms for cancer and its successful treatment. It is thus highly desirable to search for unifying concepts which would be applicable (at least) to a sufficiently large variety of tumors so as to aid in the formulation of a general phenomenological model of tumor invasion, which could then be made into a quantitative model using mathematical methods and computer simulations.

Tissue and tumor vascularization plays a critical role in determination of cellular and interstitial pH. It is well known that tumor size and metastatic potential is dependent on acquisition of adequate vasculature (Hanahan and Folkman, 1996). Tumor cell clusters that exceed 0.125 $mm^3$ must acquire vasculature or they are unable to successfully proliferate (Folkman, 1971). With regard to the current proposed model, we believe that vascularization is not only critical for nutrient exchange and tumor cell growth, but also exerts a direct effect on creation and maintenance of the acidic microenvironment postulated to favor tumor cell growth (at the expense of normal tissue).

Physiologic pH modulation in vascular beds has been well-studied. For example, the maintenance of the urea and pH gradient in the renal medulla and papilla is heavily dependent on the specialized vascular arrangement of the renal countercurrent mechanism. The presence of normal vascular beds throughout the body maintains interstitial pH, buffering large changes in pH through fluid and ion exchange, along concentration and facilitated gradients. Clearly, inadequate vascularization is associated with tissue hypoxia, cell injury, and both apoptosis and necrosis. A common observation in tumor biopsies is that areas of tumor that are poorly vascularized are most likely to be necrotic. This suggests that while the growth of tumor cells may be favored by a modestly acidic microenvironment (perhaps generated and maintained with byproducts of anaerobic glycolysis), large pH decrements associated with hypoxia will lead to the death of normal and tumor cells.

Recently, Patel *et. al.* (2001) studied a hybrid cellular automaton model of tumor invasion based on the acid-mediated invasion hypothesis of Gatenby *et.al* (1996). In Patel's model, normal and neoplastic cells and micro-vessels occupy sites of a grid array. There are also two continuously varying density fields, corresponding to acid and nutrient (glucose) concentrations. Neoplastic cells act as the source of acid in the system, while both normal and neoplastic cells act as sinks for glucose. The micro-vessels, distributed at random with a certain mean density, act as sources for glucose and sinks for the acid. In the model, growth of a tumor, starting from an initial avascular seed is then studied, using the dynamics of the acid and glucose concentration. They found that both normal and neoplastic cells become quiescent, or die at sufficiently high/low values of acid/glucose concentration. The vacant space created by the death of normal cells is filled by cancer cells through cell division. Previous numerical simulations of the model showed that, depending on the acid production rate and the mean micro-vessel density (MVD), the tumor is found to be in one of the following stages of growth: sustained growth, no growth or slow growth and growth followed by self-poisoning. However, demarcation between these different regions was not sufficiently well-established in the simulations. For example, it would be important to know the fate of a tumor with a necrotic core at its center, i.e., does the necrotic core expand throughout the tumor over sufficiently long times, or is it essentially segregated within the tumor, allowing the tumor to grow indefinitely? Such questions are difficult to answer through numerical simulations alone primarily because of the limitations imposed by finite system size, and mathematical analysis, even if only approximate, becomes an essential tool.

Our principal aim in this paper is study the correlation between micro-vessel density and the growth characteristics of a tumor in the context of the acid-mediated invasion model. The specific questions that we would like to address here are as follows: (i) What is the range of MVD suitable for growth? (ii) What are the conditions for the formation of a necrotic core at the center of the tumor? (iii) How does the necrotic core grow with the tumor? (iv) What is the effect of spatial non-uniformity of MVD (as in neo-angiogenesis) on the growth of the tumor? In order to address these questions, we use a mean-field analysis combined with cellular automaton simulations. Within the mean-field theory, we obtain the steady state profile of the acid concentration as a function of the distance from the center of the tumor, and determine the range of MVD suitable for growth of the tumor. We show that, at sufficiently low MVD, the tumor has a necrotic core at the center, which grows with the tumor. At late stages of growth, the ring of healthy tumor cells at the outer edge of the tumor has a constant width which is independent of time. We also show that having a higher MVD inside the tumor helps in increasing this width, and thus might rescue the tumor from self-poisoning. These predictions are compared to cellular automaton simulations and recent experimental results on Morris 7777 hepatoma tumor growth in Lewis rats. In general, we find good agreement on the correlation between MVD and tumor growth.

The content of this paper is arranged as follows: In
- Sec.2., features of the acid-mediated tumor growth model, along with the relevant equations,
- Sec.3, mean-field analysis of the model in the various stages of growth, starting from the initial avascular seed. We obtain expressions for the acid concentration profile as a function of the distance from the center of the tumor, as well as implicit expressions for the necrotic core radius and the radius of the hypo-cellular region, as functions of the tumor radius. We solve the steady state of these equations for several interesting regimes, and outline the various stages of growth thus identified,

- Sec.4, predictions based on the mean-field theory are tested through explicit simulations of the cellular automaton model,
- Sec.5, experimental results
- Sec. 6, summary of results and conclusions.

## 2. Acid-mediated tumor growth model

This model is based on the cellular automaton model introduced by Patel. *et. al.* (2001), summarized here. Let us consider an initial (avascular) tumor seed of radius $R_0$ placed in the center of a (two-dimensional) tissue. The tumor cells are assumed to produce acid at a rate **h** per unit time, which diffuses into the surrounding intercellular space, and produces a more acidic micro-environment for the healthy normal cells in the tissue. It is assumed that there is a sharp threshold acid concentration $\rho_N^*$, above which the normal cells would not survive. At a lower threshold, which we denote by $\rho_N^Q$, the cells would become quiescent. Micro-vessels are assumed to be present everywhere in the tissue at a certain mean surface density (denoted by $\phi$, and defined as the number of micro-vessels per cell)[1], which remove excess acid from the tissue and supply the vital nutrients for the cells. The pH level in the intercellular space increases when the removal of acid is not efficient enough, leading first to quiescence and, later, to death of normal cells. The removal of normal cells creates a 'hypocellular' gap in the tissue, which is slowly filled by neoplastic cells through cell division. This is the essential mechanism of acid mediated tumor invasion.

Excessive levels of acid are detrimental to the survival of the tumor cells as well, although they can tolerate a much higher level of acid than the normal cells. They may, in fact, be selected for their ability to grow in an acid-rich environment. Accordingly, we assume that, like normal cells, the neoplastic cells also die above a threshold acid concentration $\rho_T^*$, which is much larger than $\rho_N^*$. This is, of course, a direct consequence of the ability of the tumor to withstand much higher levels of acidity. (In principle, the tumor cells may also become quiescent beyond a sufficiently high acid concentration beyond a threshold value, which we denote by $\rho_T^Q$. However, in order to reduce the number of parameters, we simplify the situation by putting $\rho_T^* = \rho_T^Q$ ). In general, therefore, a tumor contains two kinds of cells, healthy and dead. The cluster of dead cells is usually found at the center of the tumor, and forms the necrotic core of the tumor.

The primary function of the micro-vessels is to supply nutrients to the cells, both normal and neoplastic. For simplicity, we consider only glucose (and ignore oxygen). Both normal and neoplastic cells consume glucose, at rates $k_N$ and $k_T$ respectively. Because tumor cells consume more glucose than normal cells, the glucose concentration profile may be expected to reach a minimum at the tumor center, and increases outward, assuming that the vasculature is located at the periphery of the tumor and not the center. Furthermore, the tumor cells are presumably able to survive at much lower glucose concentrations than normal cells.

---

[1] To convert to the experimentally measured number of microvessels per unit area, we need to consider the quantity $\Phi = \phi/\Delta^2$, where $\Delta$ ($\approx 20\mu$ m) is the typical linear separation between two cells in the tissue.

In the present paper, we would like to focus on the role of the low pH environment in the invasion of the tumor. Accordingly, we assume the existence of conditions such that nutrient depletion or other such mechanisms do not affect the state of tumor or normal cells significantly. We are also particularly interested in understanding the role of the micro-vessel (surface) density $\phi$ in determining the invasive character of the tumor, which we carry out using a mean-field analysis in the next section.

## 3. Mean-field analysis

In this section, we employ a mean-field approximation to study the growth characteristics of the tumor. In the first step of the calculation, we consider the initial avascular tumor seed and derive the condition for its growth. The equations for time evolution of acid concentration $\rho$ are:

$$\frac{\partial \rho}{\partial t} = D\nabla^2 \rho + h \qquad ; \qquad r \leq R_0 \qquad (1a)$$

$$\frac{\partial \rho}{\partial t} = D\nabla^2 \rho - \frac{4q\phi}{\Delta}(\rho - \rho^*) \qquad ; \quad r \geq R_0 \qquad (1b)$$

The first term in both the equations describes the diffusive dynamics of the acid field. The term h in the first equation represents the acid production by the tumor cells. The second term in eqn.(1b) represents the removal of acid from the system by the micro-vessels. The symbol q is the permeability of the micro-vessels to acid, and $\Delta$ is the typical intercellular spacing, which we approximately specify as 20 $\mu$m. This is the essential point of the mean-field approximation we are using. The presence of the micro-vessels is included through an average acid removal term proportional to the mean MVD $\phi$ (which is defined as the number of micro-vessels per cell). The serum acid concentration is denoted by $\rho^*$. The experimentally measured values of some of these parameters have been summarized in Table 1.

### 3.1 The avascular stage of growth:

The time scale of fluctuations in $\rho$ is far less than the time scale for growth of the tumor by cell division(which is of the order of many hours to days). Hence, we can use an adiabatic approximation, and assume that over time scales much less than that required for cell division, the acid density field reaches a steady state. One can then look at the steady state profile of $\rho$ by putting the time derivative in the l.h.s of the eqns.(1) and (2) to zero and solving the remaining time-independent equations with appropriate boundary conditions. We assume spherical symmetry since the initial tumor seed is circular in shape, so that $\nabla^2 \equiv \frac{1}{r}\frac{\partial}{\partial r}\left(r\frac{\partial}{\partial r}\right)$, and the solution of eqn.[1a] and [1b] is

$$\rho(r) = \rho(0) + \frac{h}{2D}r^2 \qquad r \leq R_0 \qquad (2a)$$

$$\rho(r) = \rho^* - h\frac{R_0 \xi}{D}\frac{K_0(r/\xi)}{K_1(R_0/\xi)} \qquad r \geq R_0 \qquad (2b)$$

where $\xi = \sqrt{\frac{D\Delta}{4q\phi}}$ is the length scale over which the effect of the tumor is seen, as far as acid concentration is concerned. It is useful to express this length scale in units of the cellular dimension, which we denote as $\Delta \approx 20\mu m$. After substituting the experimentally measured values of the other parameters (ref. Table I), we see that $\xi \cong \frac{3.36}{\sqrt{\phi}}\Delta$, when expressed in terms of the cellular dimension.

The condition for growth of tumor is $\rho(R_0) \geq \rho_N^*$, for which a necessary condition is

$$\phi \leq \frac{\Delta h}{4q(\rho_N^* - \rho^*)} \qquad (3)$$

The result implies that the initial avascular seed will grow further only if the mean MVD is sufficiently small. At higher values of $\phi$, the micro-vessel distribution is efficient in removing excess acid from the system, so that a condition favorable to the growth of the tumor is absent. If this condition is satisfied, then the tumor starts growing, and enters the vascular stage of growth, where micro-vessels are present inside the tumor, which could be either new vessels created through inducing neo-angiogenesis or simply the pre-existing vasculature. The vascular stage of growth will be the subject of the following two sub-sections.

**3.2 The vascular stage of growth:**

For simplicity, we start by assuming that, in this stage, the tumor has the same vascularity inside it as in the surrounding tissue. In this limit, the equations become

$$\frac{\partial \rho}{\partial t} = D\nabla^2 \rho - \frac{4q\phi}{\Delta}(\rho - \rho^*) + h \qquad r \leq R$$

$$\frac{\partial \rho}{\partial t} = D\nabla^2\rho - \frac{q\phi}{4\Delta}(\rho - \rho^*) \qquad r \geq R$$

Using the steady state condition as before, the steady state profile of $\rho$ can be found by putting the time derivatives to zero and then solving the equations:

$$\nabla^2\rho = \frac{4q\phi}{D\Delta}(\rho - \rho^*) - \frac{h}{D} \qquad r \leq R \qquad (4a)$$

$$\nabla^2\rho = \frac{4q\phi}{D\Delta}(\rho - \rho^*) \qquad r \geq R \qquad (4b)$$

The equations can be reduced to the Bessel equation in both cases after doing a suitable transformation of variables. The general solution in both cases is a linear combination of the Bessel functions $I_0(r/\xi)$ and $K_0(r/\xi)$ [1]. The coefficients are found using the conditions of continuity of the functions and their derivatives at r=R, along with the boundary condition

$$\rho(r) \to \rho^* \text{ as } r \to \infty \qquad (5a)$$

and the condition of smoothness of the solution at the origin, which requires

$$\frac{\partial \rho}{\partial r} = 0 \text{ at } r = 0. \qquad (5b)$$

The final solution satisfying these requirements is

$$\rho(r) = \rho^* + \frac{\Delta h}{4q\phi}\left[1 - \frac{R}{\xi}K_1(R/\xi)I_0(r/\xi)\right] \qquad r \leq R \qquad (6a)$$

$$\rho(r) = \rho^* + \frac{\Delta h}{4q\phi}\frac{R}{\xi}I_1(R/\xi)K_0(r/\xi) \qquad r \geq R \qquad (6b)$$

It is obvious that the acid concentration reaches a maximum at the centre of the tumor, where it has a value

$$\rho(0) = \rho^* + \frac{\Delta h}{4q\phi}[1 - \frac{R}{\xi}K_1(R/\xi)] \tag{7}$$

Since $K_1(z)$ decreases with the argument, $\rho(0)$ increases with the tumor radius R. For large tumors ($R \gg \xi$), it reaches a saturation value

$$\rho(0) \to \rho^* + \frac{\Delta h}{4q\phi} \quad \text{as} \quad \frac{R}{\xi} \to \infty$$

### 3.3 Necrotic core formation:

Let us now find the condition for formation of a necrotic core at the center of the tumor. Clearly, the necessary and sufficient condition for this is

$$\rho(0) \geq \rho_T^*$$

From the previous solution (and assuming that the tumor is sufficiently large so that $R \gg \xi$) we find that this condition becomes

$$\phi \leq \frac{\Delta h}{4q(\rho_T^* - \rho^*)} \equiv \phi_c \tag{8}$$

where $\phi_c$ is a threshold value, above which there is no necrotic region inside the tumor. For $\varphi < \varphi_c$, therefore, a necrotic core forms inside the tumor when the tumor becomes sufficiently large.

Let us now assume that there is a necrotic core with radius $R^*$ in the center of the tumor. Since the dead cells do not actively produce and secrete acid (although cell degeneration will result in acid products of hydrolysis), and it is reasonable to assume that such excessive acid levels will destroy the vascularity inside the core, the modified dynamical equations for the acid concentration are

$$\frac{\partial \rho}{\partial t} = D\nabla^2 \rho \quad ; \quad r \leq R^*$$

$$\frac{\partial \rho}{\partial t} = D\nabla^2 \rho - \frac{4q\varphi'}{\Delta}(\rho - \rho^*) + h \quad ; \quad R^* < r \leq R$$

$$\frac{\partial \rho}{\partial t} = D\nabla^2 \rho - \frac{4q\phi}{\Delta}(\rho - \rho^*) \quad ; \quad r > R$$

In the above set of equations, we have allowed for the possibility that the vascularity of the non-necrotic part of the tumor can be different from the vascularity outside the tumor. (In general, the tumor is well vascularized during angiogenesis, and hence the mean MVD inside the tumor might be significantly higher than outside). The steady state solution to this set of equations is

$$\rho(r) = \rho^*_T \qquad\qquad r < R^* \qquad (9a)$$

$$\rho(r) = \rho^* + \frac{\Delta h}{4q\varphi'} + b_1 K_0\left(\frac{r}{\xi'}\right) + b_2 I_0\left(\frac{r}{\xi'}\right) \qquad R^* \leq r < R \qquad (9b)$$

$$\rho(r) = \rho^* + b_3 K_0\left(\frac{r}{\xi}\right) \qquad\qquad r \geq R \qquad (9c)$$

where we have defined another length scale $\xi' = \sqrt{\frac{D\Delta}{4q\phi'}}$. As in the previous case, the parameters are determined using the conditions of continuity of the functions and their derivatives at $r=R^*$ and $r=R$ and the boundary conditions given by eqns.(5a) and (5b). The coefficients are thus found to be

$$b_1 = \left[\rho^*_T - \rho^* - \frac{\Delta h}{4q\phi'}\right]\frac{R^*}{\xi'} I_1\left(\frac{R^*}{\xi'}\right)$$

$$b_2 = \left[\rho^*_T - \rho^* - \frac{\Delta h}{4q\phi'}\right]\frac{R^*}{\xi'} K_1\left(\frac{R^*}{\xi'}\right)$$

and

$$b_3 = \frac{\xi}{\xi'}\left[b_1 \frac{K_1\left(\frac{R}{\xi'}\right)}{K_1\left(\frac{R}{\xi}\right)} - b_2 \frac{I_1\left(\frac{R}{\xi'}\right)}{I_1\left(\frac{R}{\xi}\right)}\right]$$

The necrotic radius $R^*$ may be implicitly expressed in terms of the tumor radius using the condition $\rho(R^*) = \rho^*_T$ which follows from the definition of $R^*$. The expression for $R^*$ turns out to be

$$K_1\left(\frac{R}{\xi}\right)\left[I_1\left(\frac{R^*}{\xi'}\right)K_0\left(\frac{R^*}{\xi'}\right)+K_1\left(\frac{R^*}{\xi'}\right)I_0\left(\frac{R^*}{\xi'}\right)+\frac{\xi'}{R^*}\frac{1}{\left[1+\frac{4q\varphi'}{\Delta h}(\rho^*-\rho_T^*)\right]}\right]=$$
$$\frac{\xi}{\xi'}\left[K_1\left(\frac{R^*}{\xi'}\right)I_1\left(\frac{R}{\xi'}\right)-I_1\left(\frac{R^*}{\xi'}\right)K_1\left(\frac{R}{\xi'}\right)\right]K_0\left(\frac{R}{\xi}\right)$$
(10)

In general, this equation may be solved numerically to find $R^*$ as a function of R, for various values of the parameters. Some of these cases will be studied in the next section. However, the previous expression may be much simplified in the limit where the tumor radius and the necrotic core radius are very large compared to the length scale $\xi$. In this limit, the limiting forms of the Bessel functions $K_{0,1}(z) \sim \frac{1}{\sqrt{2\pi z}}e^{-z}$ and $I_{0,1}(z) \sim \sqrt{\frac{\pi}{2z}}e^z$ as $z \to \infty$ may be used (Abramowitz and Stegun, 1964). The resulting expression for $R^*$ turns out to be

$$\frac{1}{1-\frac{4q\varphi'}{\Delta h}(\rho_T^*-\rho^*)}=\left[\frac{\xi}{\xi'}\sinh\left(\frac{R-R^*}{\xi'}\right)+\cosh\left(\frac{R-R^*}{\xi'}\right)\right]\sqrt{\frac{R^*}{R}} \quad (11a)$$

Let us now consider the case where the vascularity is unaffected by the tumor growth. (i.e., when $\xi'=\xi$). The simplified equation for $R^*$ is

$$\frac{1}{1-\frac{4q\varphi'}{\Delta h}(\rho_T^*-\rho^*)}=e^{\frac{R-R^*}{\xi}}\sqrt{\frac{R^*}{R}}$$

(11b)

Our aim now is to determine how $R^*$ changes with R as the tumor grows. Let us define the ratio $\eta=\frac{R^*}{R}$ and the difference $\delta=R-R^*$ between the two radii, such that $\delta=R(1-\eta)$. We now assume that $\eta \approx 1$ for large R, in which case $\delta$ is given by

$$\delta \approx -\xi\log[1-\frac{4q\varphi'}{\Delta h}(\rho_T^*-\rho^*)] > 0 \quad (11c)$$

The following conclusions may be made at this point from the mean-field analysis: The ratio of the necrotic core radius to the total tumor radius tends to unity as the tumor becomes larger, but the difference between the two radii approaches a constant as $R \to \infty$, i.e., there is a ring of healthy tumor cells at the outer edge of the tumor, whose width is constant in time. This is confirmed by numerical solution of the complete eqn.(11a), which is presented in the next section. In fact, numerical analysis shows that this property is true even when the MVD is non-uniform, i.e, when $\varphi' > \varphi$, but the ratio difference $\delta$ is enhanced by the non-uniformity. We also note that when $\varphi$ is very small, the tumor could become auto-toxic by self-poisoning with excess acid. From Eq. (11c), this would happen when $\delta$ becomes less than the typical cellular dimension, i.e., when $\delta \leq \Delta$.

It is also important to note that even if the necrotic core radius $R^*$ is less than the tumor radius R, it doesn't necessarily imply that the tumor is in a state of growth. Because, in general, there might be a ring of quiescent tumor cells surrounding the necrotic core. If the radius of this ring 'exceeds' the radius of the tumor, the tumor is unable to grow further, since the quiescent cells do not reproduce. In other words, the tumor is in a state of growth, only if the following condition is also satisfied:

$$\rho(R) < \rho_N^Q$$

These conditions will be discussed further in the next section, when we discuss the numerical solutions to the steady state equations.

**3.4 Numerical solution of the mean-field equations:**

We solved eqn.(10) numerically using the bisection method (Press et. al, 1992) in order to determine the radius of the necrotic core as a fraction of the tumor radius. Fig. 3 shows the cases of uniform vascularity and Fig.4 shows the cases where the tumor has higher vascularity inside it. We have studied MVD of 0.05, 0.1, 0.2 and 0.3, all of which are in the necrotic-core forming regime.

In Fig.3 A, we have shown the ratio difference $\delta/\xi$ as a function of $R/\xi$ for five different MVD values. We see that the difference approaches a constant value as the tumor radius increases, and this value increases with $\varphi$. When the absolute value of $\delta$ becomes too small (less than one cell dimension), the necrotic core would invade the tumor, and the tumor will self-poison with acid. For the cases studied here, $\varphi = 0.05$ and $\varphi = 0.1$ are observed to fall in this regime. In Fig. 3B, we have shown the ratio $R^*/R$ of the radii as a function of $R/\xi$. The ratio approaches unity at large values of R for all values of $\varphi$.

In Fig.4 A and B, we have shown the effect of introducing a difference in MVD inside and outside the tumor. In all the curves the vascularity outside the tumor is fixed at $\varphi = 0.1$, but the vascularity inside is varied between 0.1, 0.2 and 0.3. Fig. 4A shows ratio difference $\delta/\xi$ as a function of $R/\xi$. We note that the difference increases with increase in the tumoral vascularity, which indicates that the higher vascularity inside the tumor removes acid better, and thus leads to an increase in the number of healthy tumor cells. As in the previous case, the ratio of the radii again approaches unity for large R in all the cases. These results show that the number of healthy cells inside the tumor can be increased in a more efficient way by increasing vascularity inside (or in the vicinity of) the tumor, rather than more uniformly.

To conclude this section, we have shown that a difference in vascularity between the tumor and the surrounding environment can rescue the tumor from self-poisoning due to excess acid. The result has obvious implications to the process of angiogenesis, where the tumor induces the creation of a highly non-uniform vasculature, which is denser as one gets closer to the tumor.

In the next section, we test our predictions based on mean-field analysis to explicit cellular automaton simulations of the model for cases of uniform as well as non-uniform vascularity.

## 4. Cellular automaton simulations

### 4.1 Hybrid Cellular automaton model

We performed cellular automaton simulations of the system using a slightly modified version of the model proposed by Patel et. al. (1996), where the tissue is imagined as a grid of automaton elements occupied by cells, normal or cancerous, and micro-vessels. The linear size of an element is fixed at $\Delta \approx 20\mu m$, which corresponds to the linear dimension of a real cell. In our model, we have distributed the cells and micro-vessels in two different grid arrays, which interact only through exchange of acid (One could also imagine the micro-vessels as occupying the space between the cells). This modification enabled us to study the effects of multiplication of blood vessels inside the tumor without increasing the system size.

The dynamics of the model proceeds according to the principles laid out in Sec.2. For the details of the simulation algorithm, we refer the reader to Patel. *et al*. (2001). The simulations were done with a square grid of linear size L=100, and were run up to 36 generations of growth. By this time, finite size effects were already visible, so any increase in the number of generations would require a larger system size, which, at present, is beyond our computational resources.

The initial condition of the system may be described as follows. Micro-vessels are distributed at random in its array with mean density $\phi$, which we vary between 0.05 and 0.5. The cell-array is initially filled with normal cells. Now a tumor seed of radius $R_0$ is placed at the center of the array, displacing the normal cells that were already occupying these points.

The dynamics of the acid field is described by the following discretized equation:

$$\rho(r, t + \Delta t) = \rho(r, t) + \frac{D\Delta t}{\Delta^2}[\Sigma \rho(r^{'}, t) - 4\rho(r,t)] + H(r)\Delta t - \frac{q\Delta t}{\Delta}[\rho(r,t) - \rho^*]\Phi(r) \quad (12a)$$

The sum in the previous equation is over all points r′ which are the nearest neighbors of the point r in the square lattice, and $\Delta t$ is the time increment in the simulations. The functions H(r) and $\Phi$(r) are defined as follows:

H(r)=h  if the point r is occupied by a healthy tumor cell and zero otherwise.

$\Phi$(r)=1  if there is a micro-vessel at the point corresponding to r in its array, and zero otherwise.

The numerical values of the various parameters used in the simulations are summarized in Table 1. As explained in Sec.2, normal (and tumor) cells become quiescent or die when the acid concentration at their location increases beyond certain threshold levels. Since the normal cells die at lower acid levels than tumor cells, a hypo-cellular gap is created at the tumor-host interface, which is filled by the tumor through cell division, which is assumed to take place every 100 hours.

The time increment $\Delta t$ is chosen such that the maximum of the following dimensionless parameter is less than a threshold value (which is required for stability of the equations 12 a and b), which we choose as below:

$$\text{Max}[\frac{D\Delta t}{\Delta^2}, \frac{q\Delta t}{\Delta}]=1/4$$

After substituting the numerical values from Table 1, we find that this condition gives $\Delta t$ =0.011 sec. We note that this time scale is several hundreds of orders of magnitude smaller than the cell division time. This vast separation of time scales is the basis for the adiabatic approximation used in the mean-field analysis.

A similar approximation may be used in the numerical simulations also. We note that the number of Monte Carlo time steps for every event of cell division is nearly $3\times10^7$. For the system size used in simulations (L=100), the number of Monte Carlo time steps required for 'equilibration' (i.e., to achieve the steady state) of a diffusive density field is of the order of $10L^2$. Thus, rather than simulating $3\times10^7$ Monte Carlo time steps, it is enough to let the acid field 'equilibrate' to its steady state concentration profile by taking the system through ~ $10L^2$ MC steps for each cell division cycle. By monitoring the spatially averaged acid concentration as a function of time, we have ensured that this time scale is longer than sufficient for 'equilibration' of the diffusing acid field.

**4.2 Cellular automaton results**

The results of numerical simulations are shown in Figs.5-7. The critical MVD $\phi_c$ defined in (8) is found to be nearly 0.42 for the parameter values used in these simulations. In Fig.5, where $\phi = 0.05$, we see that the tumor has undergone self-poisoning after four generations. On the other hand, when a higher MVD (=0.4) is used, the tumor is found to be in a state of growth even after 36 generations (Fig.6), although some necrotic cells could be seen at the center.

In 7A, we show the system after 36 generations, when $\phi = 0.2$. Although self-poisoning has not occurred yet, we observe that the necrotic core at the center occupies a sizeable fraction of the total tumor area. In Fig.7B, we have introduced non-uniformity in the vascularity by having a higher MVD inside the tumor ($\phi' \approx 0.4$). This is implemented in the simulations in the following way. At each generation, the micro-vessels inside the tumor would 'reproduce' once with probability p, and the newly created daughter vessel is then placed at a random location inside the tumor which is not already occupied by a micro-vessel. By suitably tagging the micro-vessels, we ensure that each micro-vessel reproduces only once. Then, by tuning the probability p, we are able to increase the MVD inside the tumor to any desired level (for example, when p=1, $\phi' = 2\phi$). Fig. 7B shows the system after 36 generations. The fraction of area inside the tumor occupied by the necrotic region is considerably smaller than the previous case.

In Fig.8, we have plotted the fraction of area occupied by necrotic cells against the number of generations both for uniform ($\phi = 0.2$) and non-uniform ($\phi' \approx 0.3$ and 0.4) MVD. In the first case, the fraction is increasing continuously with the number of generations (and will asymptotically reach 1 according to mean-field predictions). However, in the simulations, finite lattice size appears to cause a bending of this curve at late times. When the MVD is higher inside the tumor compared to outside, the fraction of area covered by necrotic region increases initially (although much more slowly compared to the previous case), but seems to approach a limiting value less than unity. However, it must also be noted that finite size effects are serious at late times, so a confirmation of this result has to come from much larger lattice sizes.

## 5. Experimental results

To provide a means to validate some of the concepts of the current study, an inducible tumor model was studied. The Morris 7777 hepatoma is a transplantable tumor originally derived from the liver of Lewis rats exposed to 2-acetylaminofluorene. This tumor is maintained both in tissue culture and frozen state and can be readily transplanted to compatible rats to study progressive tumor growth. As noted, this is an hepatocellular carcinoma, which proliferates first as a discrete avascular cell cluster and later as in infiltrative tumor within tissue.

Groups of anesthetized female Lewis rats were injected with a cell suspension of $10^6$ tumor cells in the fat pad medial to the right thigh. Tumor growth was monitored by palpation and measurement daily for a period of nine weeks. At selected weekly intervals, rats were humanely sacrificed and carefully necropsied. Gross lesions were recorded, including tumor dimensions and growth characteristics. Sections of tumors were collected by dissection, fixed in 10% neutral buffered formalin, processed, and sectioned. Hematoxylin-eosin stained sections were examined by an experience veterinary pathologist, and these sections were photographed. For analysis, the center of tumors was determined by measurement, and then photomicrographs taken every 1 mm from the center of the tumor to the boundary with normal, non-tumor tissue. In a selected number of tumors, the percentage of viable and necrotic tumor cells was determined subjectively and an assessment was made of the amount and type of vascularization was made. Three tumors selected for analysis (#6, #12, and #40) represented early (2 weeks post implantation), middle (3 weeks post implantation) and late (9 weeks post implantation) stages of tumor growth. The results of observations are tabulated in Tables 2 and 3.

Several observations, relevant to the present mathematical model, can be made from these observations. First, with progressive growth and enlargement of the tumor, tumor cell viability was affected. Tumor #6 had a measured cross sectional area of approximately 1 cm$^2$, while tumor #12 was approximately 2 cm$^2$, and tumor #40 approximately 4cm$^2$. Virtually none of tumor #6 was necrotic, but areas and bands of necrotic tumor tissue were present in other tumors, which were larger in size. This is in very good agreement with the theoretical predictions made in Sec. 3.2. Second, areas of necrosis were relatively devoid of vasculature, and presumably were acidic, as the result of proteolysis and cell degradation.

## 6. Discussion and Conclusions

In this paper, we have presented a study of the various stages of growth of a tumor, where the invasion mechanism is the acidification of the micro-environment around it. We made use of the vast difference in the time scales of tumor growth, and the equilibration of the acid concentration profile, in order to employ an 'adiabatic' approximation. In this approximation, the tumor radius is a slowly varying variable, and is essentially treated as a parameter in terms of which other variables and fields are expressed. This approach is combined with a mean-field approximation and the steady state profile of the local acid concentration is determined as a function of the distance from the radius of the tumor.

Using the steady state profile, we determined the conditions for various stages of tumor growth. In particular, when the micro-vessel density is assumed to be uniform everywhere, the mean-field approximation predicts the existence of three regimes of growth: growth followed by auto-toxicity, sustained growth and no growth. Introducing a difference between the MVD inside and outside the tumor (as would occur in angiogenesis) was predicted to rescue the tumor from self-poisoning, and, for sufficiently large difference, would help the tumor to grow. These predictions were supported by explicit cellular automaton simulations.

The mean-field analysis presented here could be easily adapted to treat other similar features of cancer growth, like nutrient deprivation. It would be interesting to see how the predictions based on the acid-mediated invasion mechanism would be affected by this modification. In particular, it would be interesting to see if the possible 'invasion' of a tumor by its necrotic core would be present if glucose deprivation, instead of low pH, were the dominant mechanism of invasion. We are currently studying this case, and expect to report our results in due time.

The experimental data show that in most instances where there is good tumor cell viability, there is a small, but adequate number of vessels supporting tumor growth. Third, there is a relatively abrupt interface between tumor tissue and surrounding tissue. In these tumors, there is formation both of a connective tissue capsule and inflammatory cell layer between fat (the site of implantation) and tumor. In terms of this model, one would expect to see altered (acidic) pH at the interface, and highly acidic pH in areas that are necrotic. Direct measurement of acid phosphatases and metalloproteinases in these areas would substantiate the hypothesis; such enzyme histochemical stains are currently being pursued.

## Ackowledgements

We gratefully acknowledge the Carilion Biomedical Institute for partial financial support for this project. The authors would like to thank Kimberly Forsten-Williams for valuable suggestions and Kenneth Meissner for fruitful discussions and suggestions throughout the course of this work. B.G would also like to thank Manoj Gopalakrishnan for helpful suggestions for the simulations.

# Figures

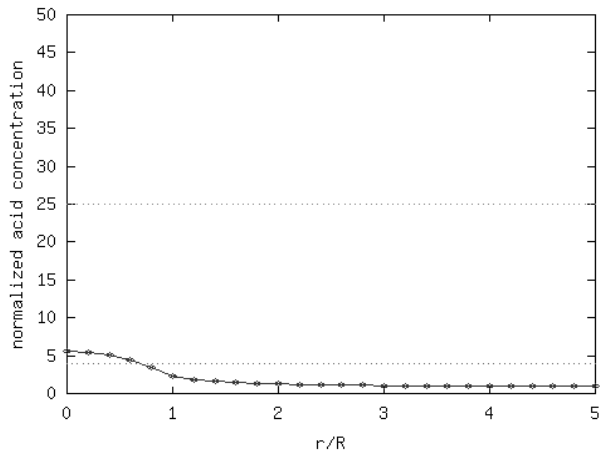

**Fig.1** The acid concentration profile (scaled using serum acid concentration) as a function of the distance from the center of the tumor (scaled with tumor size). The top horizontal line is the level for death of cancer cells, and the bottom horizontal line is the corresponding level for normal cells. The MVD level is 0.9 here, much higher than the threshold value for cancer growth.

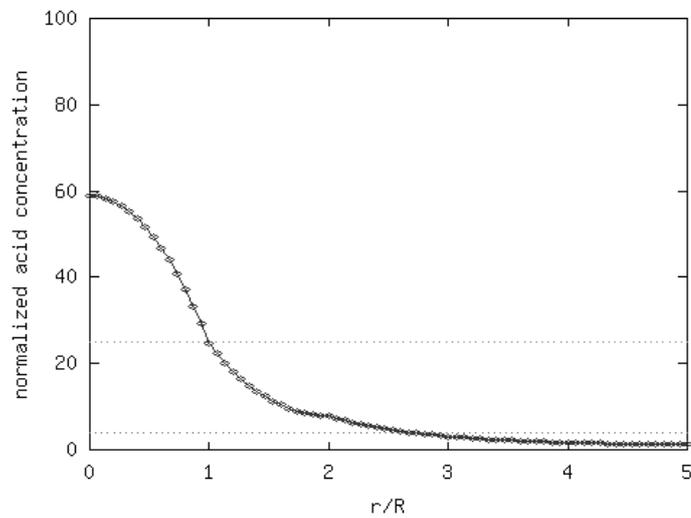

**Fig.2** Same as Fig.1, but for a lower value of MVD (=0.05). In this regime, the tumor has undergone self-poisoning. The threshold MVD for auto-toxicity is nearly 0.42 for the parameters used here (Ref: Table I).

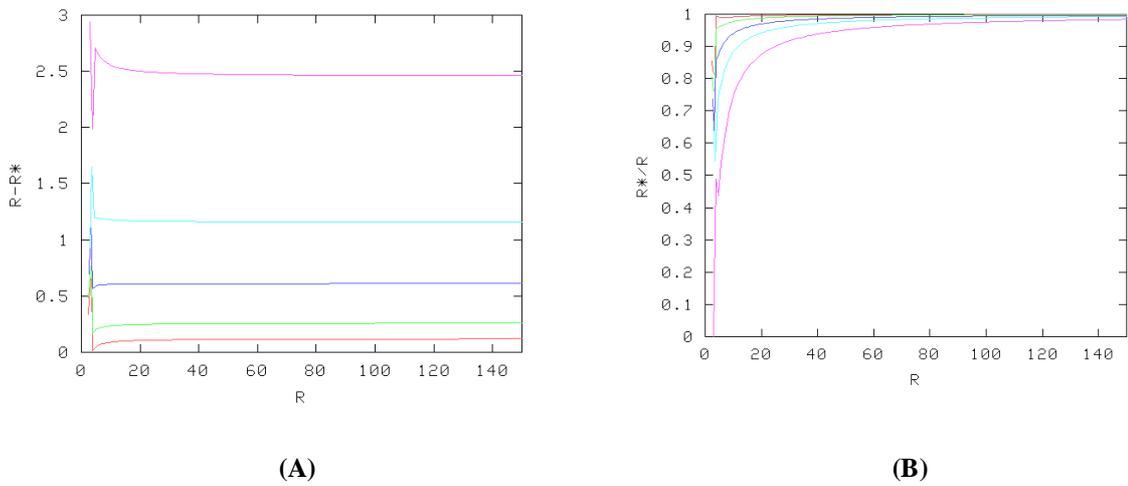

**(A)**                 **(B)**

**Fig.3** The figure shows the growth of necrotic core inside the tumor with increase in tumor radius. We have plotted here (A) The difference in radii between the necrotic core and the whole tumor, and (B) the ratio between the radii, as a function of the tumor radius. Both x and y axes have been scaled by $\xi$. The curves correspond to $\varphi$ = 0.05, 0.1, 0.2, 0.3 and 0.4 (Bottom to Top). These plots have been obtained by numerical solution of eqn(10), and the various parameter values are given in Table I.

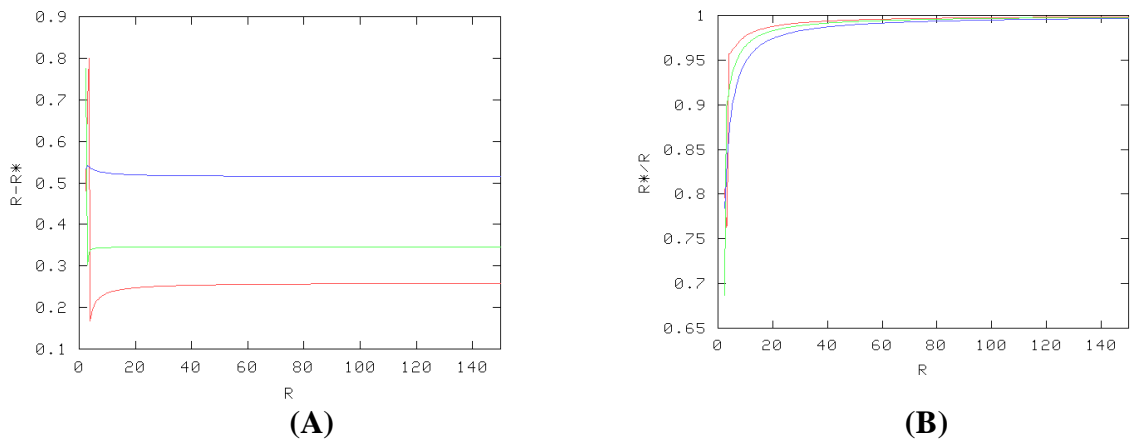

**(A)**                 **(B)**

**Fig.4** The figure shows the growth of necrotic core inside the tumor with increase in tumor radius, when the MVD inside the tumor is higher than outside. We have plotted here (A) The difference in radii between the necrotic core and the whole tumor, and (B) the ratio between the radii, as a function of the tumor radius. Both x and y axes have been scaled by $\xi$. The curves correspond to $\varphi'$ = 0.1, 0.2 and 0.3. (Bottom to Top), and $\varphi$ =0.1 always. These plots have been obtained by numerical solution of eqn(10), and the various parameter values are given in Table I. Note that higher MVD inside the tumor leads to an increase in the number of healthy tumor cells, thus making the tumor more invasive.

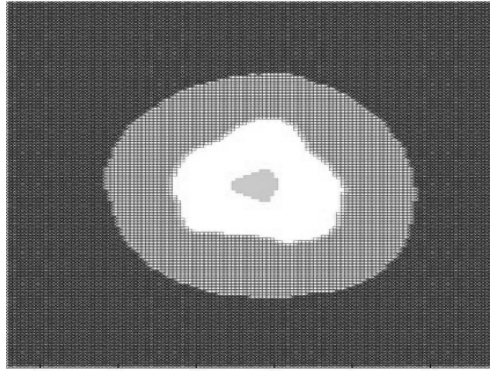

**Fig.5.** The figure shows the result of cellular automaton simulation of the model at $\phi$ =0.05 (in the necrotic core forming regime). The tumor has undergone self-poisoning here (the dark portion inside corresponds to necrotic cells), in agreement with the mean-field predictions. The grey ring represents quiescent normal cells and the dark background is the healthy normal tissue. In this, as well as following figures, the white region is the hypo-cellular gap between tumor and normal cells.

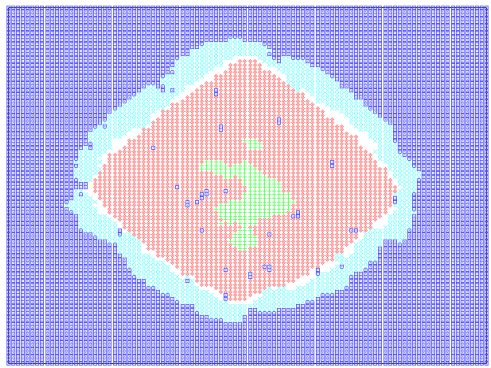

**Fig.6**. In this case ($\phi = 0.4$), the tumor is still growing after 36 generations, with a small necrotic region at the center. The lattice size used here is L=100. The island inside the tumor is the necrotic core, and the ring surrounding the tumor represents quiescent normal cells. Tumor quiescence is suppressed here The blue sea outside is the healthy normal tissue.

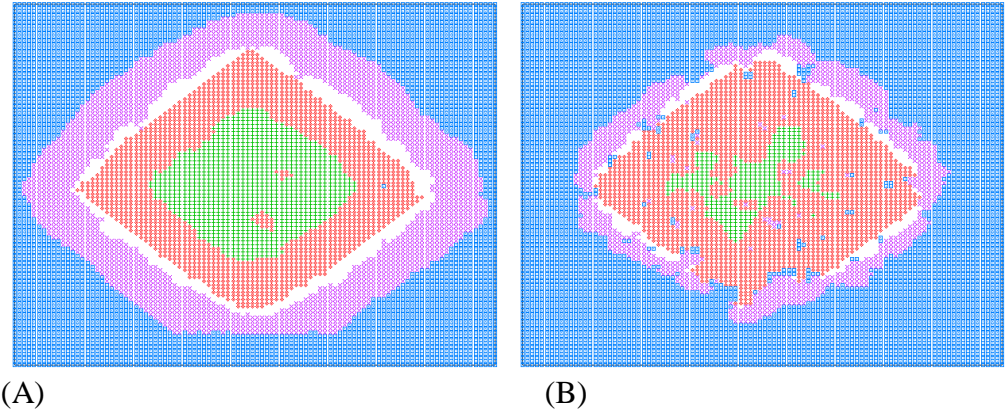

(A)                              (B)

**Fig. 7.** Cancer growth after 36 generations when (A) $\phi = 0.2$ both inside and outside the tumor, and (B) $\phi = 0.2$ (outside) and $\phi' = 0.4$ (inside). The green portions inside the tumor are the necrotic cells. We observe that increasing the MVD inside the tumor has restricted the size of the necrotic region. Tumor quiescence is suppressed here. The lattice size used is L=100, and the pictures were taken after 36 generations. Note that the rate of growth in these cases is higher than the previous case (Fig.6).

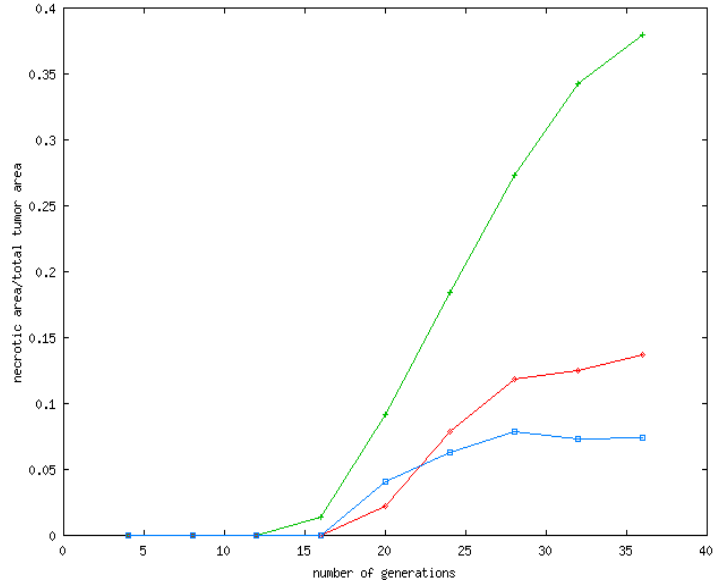

**Fig. 8**. The area of the necrotic region as a fraction of the total tumor area is plotted as a function of the number of generations when (Top) $\phi = 0.2$ both inside and outside the tumor, (Middle) $\phi = 0.2$ (outside) and $\phi' = 0.4$ (inside), and (Bottom) $\phi = 0.2$ (outside) and $\phi' = 0.3$ (inside). In the uniform case, the ratio keeps increasing with the number of generations, whereas in the other two cases, it appears to saturate, presumably due to finite size effects.

| | |
|---|---|
| D | $1.08 \times 10^{-5}$ cm$^2$ s$^{-1}$ |
| q | $1.19 \times 10^{-4}$ cms$^{-1}$ |
| $\Delta$ | $2.0 \times 10^{-3}$ cm |
| $\Delta t$ | 0.0109 s |
| $\rho^*$ | $3.98 \times 10^{-5}$ mM |
| $\rho_\Gamma^*$ | $1.0 \times 10^{-3}$ mM |
| $\rho_N^*$ | $1.585 \times 10^{-4}$ mM |
| $\rho_N^Q$ | $7.94 \times 10^{-5}$ mM |
| h | $1.0 \times 10^{-4}$ mM s$^{-1}$ |
| $\phi$ | 0.05-0.4 ($\approx 10^4 - 10^5$ cm$^{-2}$) |

**Table 1.** The numerical values of the various parameters used in the simulations, as obtained from Patel et. al (2001).

| Picture | V % | N % | SmV | MedV | LgV | CT | Inf |
|---|---|---|---|---|---|---|---|
| 6/1 | 100 | 0 | 1 | - | - | - | - |
| 6/2 | 95 | 5 | 1 | 1 | - | - | - |
| 6/3 | 95 | 5 | 3 | 3 | - | 1 | - |
| 6/4 | 95 | 5 | 1 | - | - | 1 | - |
| 6/5 | 95 | 5 | 1 | - | - | - | - |
| 6/6 | 100 | 0 | 1 | - | - | 2 | 3 |
| 6/7 | Fat | | | | | | |
| 6/8 | Fat | | | | | | |
| 6/9 | Fat | | | | | | |
| 6/10 | Fat | | | | | | |
| 6/11 | Muscle | | | | | | |
| | | | | | | | |
| 12/1 | 5 | 95 | - | - | - | 1 | - |
| 12/2 | 70 | 30 | 1 | - | - | 1 | - |
| 12/3 | 90 | 10 | 1 | - | - | 2 | - |
| 12/4 | 80 | 20 | 2 | - | - | 1 | - |
| 12/5 | 70 | 30 | 2 | - | - | - | - |
| 12/6 | 40 | 60 | 1 | 1 | - | 1 | - |
| 12/7 | 90 | 10 | 1 | 1 | - | 1 | - |
| 12/8 | 95 | 5 | 1 | 2 | - | 2 | 3 |
| 12/9 | Fat | | | | | | |
| 12/10 | Muscle | | | | | | |

**Table 2: Results of observations of tumor growth. Rats #6 and #12.**

**Abbreviations:**
V % - percentage of viable tumor cells in field
N % - percentage of necrotic tumor cells in field
Density of vessels:
SmV- small vessels (capillaries, sinusoid-like)
MedV- medium vessels (small venules)
LgV – large vessels (medium venules)
        1-few, scattered
        2-small number, scattered
        3- moderate number, scattered
        4- large number, scattered
CT-fibrous connective tissue (1-minimal, 2-moderate, 3-marked)
Inf- inflammatory cells (1-minimal, 2-mild, 3-moderate, 4-marked)

|       | V %    | N %  | SmV | MedV | LgV | CT | Inf |
|-------|--------|------|-----|------|-----|----|-----|
| 40/11 | 70     | 30   | 1   | 1    | -   | -  | -   |
| 40/12 | 60     | 40   | 1   | -    | -   | 1  | -   |
| 40/13 | 80     | 20   | 1   | -    | -   | 1  | -   |
| 40/14 |        |      | 1   | -    | 1   | 2  | 4   |
| 40/15 | 50     | 50   | 2   | -    | -   | 1  | 3   |
| 40/16 | 50     | 50   | 1   | -    | -   | 2  | -   |
| 40/17 | 80     | 20   | 1   | -    | -   | 1  | -   |
| 40/18 | 90     | 10   | 1   | 2    | -   | 1  | -   |
| 40/19 | 70     | 30   | 1   | -    | -   | 1  | -   |
| 40/20 | 95     | 5    | 2   | 1    | -   | -  | -   |
| 40/21 |        |      | 2   | 1    | -   | 1  | 4   |
| 40/22 | 95     | 5    | 1   | -    | -   | 1  | -   |
| 40/23 | 90     | 10   | 1   | -    | -   | 1  | -   |
| 40/24 | 95     | 5    | 2   | -    | -   | -  | -   |
| 40/25 | 95     | 5    | 1   | -    | -   | 2  | -   |
| 40/26 | 95     | 5    | 1   | -    | -   | 1  | -   |
| 40/27 | 80     | 20   | 1   | 2    | -   | 2  | 3   |
| 40/28 | 80     | 20   | 1   | 2    | -   | 3  | 3   |
| 40/29 | Fat    |      |     |      |     |    |     |
| 40/30 | Fat    |      |     |      |     |    |     |
| 40/31 | Muscle |      |     |      |     |    |     |
| 40/32 | Muscle |      |     |      |     |    |     |

**Table 3: Results of observations of tumor growth. Rat #40.**

**Abbreviations:**
V % - percentage of viable tumor cells in field
N % - percentage of necrotic tumor cells in field
Density of vessels:
SmV- small vessels (capillaries, sinusoid-like)
MedV- medium vessels (small venules)
LgV – large vessels (medium venules)
        1-few, scattered
        2-small number, scattered
        3- moderate number, scattered
        4- large number, scattered
CT-fibrous connective tissue (1-minimal, 2-moderate, 3-marked)
Inf- inflammatory cells (1-minimal, 2-mild, 3-moderate, 4-marked)